\documentclass[aps,prl,reprint,superscriptaddress,showpacs,dvipdfmx]{revtex4-1}
\usepackage{graphicx}
\usepackage{bm}
\usepackage{color}

\begin{document}

\title{
Generic Weyl phase in the vortex state of quasi-two-dimensional chiral superconductors
}

\author{Tomohiro Yoshida}
\affiliation{Department of Physics, Gakushuin University, Tokyo 171-8588, Japan}
\author{Masafumi Udagawa}
\affiliation{Department of Physics, Gakushuin University, Tokyo 171-8588, Japan}

\date{\today}

\renewcommand{\k}{{\bm k}}

\begin{abstract}  
  We study the collective behavior of Majorana modes in the vortex state of 
 chiral $p$-wave superconductors.
   Away from the isolated vortex limit, the zero-energy Majorana states communicate with each other on a vortex lattice,
and form a coherent band structure with non-trivial topological character.
We revealed that the topological nature of Majorana bands changes sensitively via quantum phase transitions in the two-dimensional (2D) systems, 
as sweeping magnetic field or Fermi energy.
  Through the dimensional reduction, we showed the existence
of generic superconducting  
Weyl phase in a low magnetic field region of quasi-2D-chiral superconductors. 
\end{abstract}


\maketitle
Topological superconductivity is one of the 
central 
topics in modern condensed matter physics.~\cite{PRB.61.10267,PRL.86.268,PU.44.131,PRB.78.195125,PRL.100.096407,PRB.79.060505,PRL.102.187001,PRB.79.094504,PRL.103.020401,PRB.82.134521,PRB.81.220504,PRL.104.040502,*PRB.81.125318,*PRL.105.077001,PRL.105.097001,RMP.83.1057,JPSJ.81.011013,Fu2014}. 
A fascinating feature of topological superconductivity is the fully gapped bulk spectrum accompanied by
topologically protected gapless boundary states. 
As a prototypical example, 
a zero-energy Majorana state appears in an isolated vortex core of spinless chiral $p$-wave superconductors~\cite{JETP.70.609,PRB.61.10267,PRL.86.268}.

  While the existence of stable zero-energy Majorana mode is striking in itself, however, its non-trivial statistical properties, as a non-Abelian anyon, show up only in multi-particle systems.
For example, the manipulation of qubits for topological quantum computation requires an entanglement of two or a larger number of particles. In this light, it is desirable to elucidate the collective behavior of Majorana modes.

  The interaction between Majorana modes, however, introduces a new problem. Namely, the zero-energy property may not be protected in the presence
of other particles. In fact, this problem has been examined by several groups~\cite{PhysRevLett.103.107001,PhysRevB.82.094504,PhysRevB.86.075115,PhysRevB.88.064514,PhysRevLett.111.136401,PhysRevB.92.134519,PhysRevB.92.134520} in chiral superconductors. 
They focus on the effect of inter-vortex tunnelings, and show that it tends to perturb the Majorana modes off the zero energy. 
This fragility of zero-energy state may be disappointing in terms of the application to e.g. topological quantum computation, however, the relevance of
interaction implies a possibility of novel cooperative phenomena inherent in Majorana many-body systems. 

In this work, we focus on the coherent band formation in the vortex state of chiral superconductors. 
In this context, Majorana modes in each vortex cores play a role of atomic orbitals in the band formation in a crystal solid. 
In this light, these Majorana modes have several fascinating properties absent in normal solids. 
Firstly, the fundamental degrees of freedom obey Majorana, rather than Fermi, commutation relations. This makes difference in the symmetry classification
of resultant band structure. Secondly, the (magnetic) unit cell contains two vortices, since each superconducting vortex carries flux $\pi$. 
Accordingly, the theoretical description of Majorana bands needs doubly enlarged magnetic unit cell with a coupling to gauge field,
which results in a fertile possibility of metal-insulator transition with topologically non-trivial character.
And thirdly, the band parameters are easily tuned by magnetic field and electron density. 
As we discuss later, these features lead to the existence of successive quantum phase transition in two-dimensional (2D) chiral superconductors.

Furthermore, this ubiquity of quantum phase transitions in 2D systems is connected to a topological property in higher-dimensional systems.
Topological phenomena in different dimensions are sometimes closely related with each other.
Through the idea of dimensional reduction, we show the existence of generic superconducting Weyl phase in 
quasi-2D chiral superconductors. 

In order to explore the topological aspect of chiral $p$-wave superconductors in the vortex state,
we start with the attractive extended Hubbard model on square and layered square lattices, whose Hamiltonian is given by 
\begin{eqnarray}
  {\cal H}&=&\sum_{\bm i}\sum_{\bm\delta}t_{{\bm i}, {\bm i}+{\bm\delta}}c^\dagger_{{\bm i}}c_{{\bm i}+{\bm\delta}}-\mu\sum_{\bm i}c_{\bm i}^\dagger c_{\bm i}
  +\frac{1}{2}\sum_{{\bm i},{\bm j}}V_{{\bm i}{\bm j}}c_{{\bm i}}^\dagger c_{\bm j}^\dagger c_{\bm j}c_{\bm i}. \nonumber \\
  \label{eq1}
\end{eqnarray}
For simplicity, we consider spinless fermion, and define 
$c_{\bm i}$ as its annihilation operator at site ${\bm i}$.
For the moment, we consider only the square lattice, and set its coordinate as ${\bm i}=i_x{\bm x} + i_y{\bm y}$. 
Here, we set the lattice space $a=1$, and define ${\bm x}$ (${\bm y}$) as a unit vector in $x$ ($y$) direction. 
The summation over ${\bm\delta}$ is taken for the vectors connecting nearest-neighbor sites: ${\bm\delta}=\pm{\bm x}, \pm{\bm y}$. 

We apply magnetic field ${\mathbf H}$ in $z$ direction, and adopt the Landau gauge for the vector potential ${\bm A}({\bm r})=(0,Hx,0)$.
Here, we assume the limit of type-II superconductivity, and ignore internal magnetic fields.
We incorporate the effect of magnetic field in the hopping term with Peierls substitution: $t_{{\bm i}, {\bm i}+{\bm x}}=-t$ and $t_{{\bm i}, {\bm i}+{\bm y}}=-te^{-i\frac{2\pi}{\phi_0}Hi_x}$, with $\phi_0=hc/e$, a flux quantum.

We analyze the model with
the Bogoliubov-de Gennes (BdG) equation:
\begin{eqnarray}
  \sum_{\bm j}\left(
  \begin{array}{cc}
    H_{\bm ij} & \Delta_{\bm ij} \\
    \Delta_{\bm ij}^\dagger & -H_{\bm ij}^\ast
  \end{array}
  \right)
  \left(
  \begin{array}{c}
    u^\nu(\bm j) \\
    v^\nu(\bm j)
  \end{array}
  \right)
  =E_\nu\left(
  \begin{array}{c}
    u^\nu({\bm i}) \\
    v^\nu({\bm i})
  \end{array}
  \right),
  \label{eq3}
\end{eqnarray}
with $H_{{\bm i}{\bm j}}=\sum_{\bm\delta}t_{{\bm i},{\bm j}}\delta_{{\bm i}+{\bm\delta},{\bm j}}-\mu\delta_{{\bm i},{\bm j}}$. 
Here, we define the order parameter: 
$\Delta_{{\bm i}{\bm j}}=V_{{\bm i}{\bm j}}\langle c_{\bm j}c_{\bm i}\rangle$, 
with $\langle\cdots\rangle$, a thermal average at the temperature, $T=0.001t$.

We assume that superconducting vortices form a square lattice. 
Since each vortex carries a flux $\pi$, a magnetic unit cell must contain two vortices.
Accordingly, we set the magnetic unit cell of dimension: 
$N_x\times N_y=N\times 2N$, corresponding to a magnetic field $H=\phi_0/N^2$.
We Fourier transform the BdG equation (\ref{eq3}), with magnetic wave vectors ${\mathbf k}=(k_x, k_y)$: $k_{x,y}\in (-\pi/N_{x,y},\pi/N_{x,y}]$,
by imposing a boundary condition on the Fourier components of $u^\nu({\bm i})$ and $u^\nu({\bm i})$, to fit with the non-periodic spatial variation of ${\mathbf A}({\mathbf r})$.

  As to the interaction, we assume nearest-neighbor attraction: $V_{\bm ij}=-V_{\rm p}\sum_{\bm\delta}\delta_{{\bm i}+{\bm\delta},{\bm j}}$,
  which selects the chiral $p$-wave superconducting state, among the five irreducible representations in the point group $D_{4h}$~\cite{RevModPhys.63.239},
  We also consider the on-site attractive interaction: 
  $V_{\bm ij}=-V_{\rm s}\delta_{{\bm i},{\bm j}}$, 
  for the $s$-wave state as a reference.

We take $t=1$ as a unit of energy, and we typically choose $\mu=-2$, which yields the electron filling: $n\sim0.18$. 
To facilitate the comparison, 
we assume slightly different values for the magnitudes of attractive interactions: 
$V_{\rm s}=2.5$, and $V_{\rm p}=2.4$,
  that yield rather close values for critical temperatures: 
$T_{\rm c}=0.17$ for $s$-wave and $T_{\rm c}=0.15$ for $p$-wave pairings.

\begin{figure}[tb]  
    \includegraphics[width=90mm]{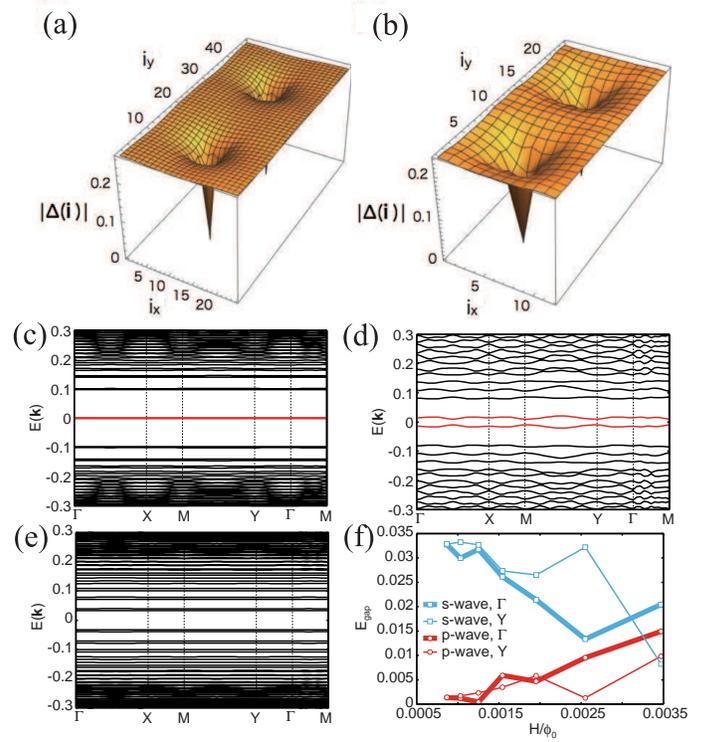}   
  \caption{(Color online) Spatial profile of the $p_x$-$ip_y$-wave order parameter at the magnetic field
    (a) $H=H_{\rm l}=8.68\times10^{-4}\phi_0$ and (b) $H=H_{\rm h}=3.47\times10^{-3}\phi_0$, and the temperature $T=0.001$.
    Coresponding energy spectra are shown in (c) and (d).
    We show the energy spectrum near the Fermi energy along the symmetric points,
    $\Gamma$ [${\bm k}=(0,0)$] $\rightarrow$ X [${\bm k}=(\pi/N_x,0)$] $\rightarrow$ M [${\bm k}=(\pi/N_x,\pi/N_y)$] $\rightarrow$
    Y [${\bm k}=(0,\pi/N_y)$] $\rightarrow$ $\Gamma$ $\rightarrow$ M.    
    For comparison, we show the energy spectrum for $s$-wave order parameter in (e) at the low magnetic field $H_l$.
    (f) Magnetic field dependence of the energy gap $E_{\rm gap}$ for 
    $s$-wave and $p$-wave superconductors at $\Gamma$ and Y point.
  }
  \label{fig1}
\end{figure}
First, we show the spatial profiles of the order parameter. Of the two possible chiral pairings, the $p_x-ip_y$ state is favored
under a magnetic field, whose order parameter $\Delta({\bm i})$ is defined as 
$\Delta({\bm i})=(\Delta_{{\bm i},{\bm i}+{\bm x}}-\Delta_{{\bm i},{\bm i}-{\bm x}}-i\Delta_{{\bm i},{\bm i}+{\bm y}}+i\Delta_{{\bm i},{\bm i}-{\bm y}})/2$.
  We plot $\Delta({\bm i})$ for $H_{\rm l}/\phi_0=8.68\times10^{-4}$ and $H_{\rm h}/\phi_0=3.47\times10^{-3}$ in Fig.~\ref{fig1}(a) and (b), respectively. 
The amplitudes of $\Delta({\bm i})$ have a small difference between these two fields.
Meanwhile, the spatial variation of $\Delta({\bm i})$ differs, considerably. Reflecting the smaller inter-vortex distance, $\Delta({\bm i})$ 
is reduced even near the boundary of magnetic unit cell at the higher field, $H_{\rm h}$.

The difference of spatial variation affects the fermionic energy spectrum of the system, considerably.
We plot the energy spectra at $H=H_{\rm l}$ and $H_{\rm h}$ in Fig.~\ref{fig1}(c) and (d), respectively. 
For $H=H_{\rm l}$, we find low-energy states near the Fermi energy, isolated from the other high energy bands. 
The low-energy states are composed of nearly degenerate two flat modes, which are separated by a small gap $\sim0.002$.
  These states are remnants of zero-energy Majorana states in the isolated chiral vortex cores\cite{JETP.70.609,PRB.61.10267}. 

While the states are exactly degenerate at Fermi energy at the limit of isolated vortices, however, 
the inter-vortex tunneling introduced the slight separation.
This tendency becomes conspicuous as increasing magnetic field, where closely-spaced vortices allow larger tunnelings.
At $H=H_{\rm h}$, the separation between the two low-energy bands is clearer [Fig.~\ref{fig1}(d)], reflecting the
larger spatial variation of $\Delta({\bm i})$. 

To compare the effects of lattice formation with a conventional case, 
we show the energy spectrum 
for the $s$-wave order parameter at $H=H_{\rm l}$ as a reference, in Fig.~\ref{fig1} (e). 
The flatness of the bands imply that there are little communications between vortices in this magnetic field. Nevertheless,
there is a clear offset from zero energy in the energy bands closest to the Fermi energy, in sharp contrast to the $p_x$-$ip_y$ case [Fig.~\ref{fig1}(c)].

In fact, as magnetic field varies, the energy gaps develop in a contrastive way between $p_x$-$ip_y$ and $s$-wave superconductors.
Figure \ref{fig1}(f) shows the magnetic field dependence of the energy gap $E_{\rm gap}$, defined as the separation of two band energies closest to the Fermi energy.
Overall, $E_{\rm gap}$ decreases (increases) as lowering magnetic field for $p_x$-$ip_y$ ($s$-wave) superconductors.
This contrastive behavior can be ascribed to the different energy level structures in the isolated vortex limit. 
In the $s$-wave case, there is already a finite gap from Fermi energy at this limit. So, the lattice formation reduces this gap by a band width.
In contrast, in the $p_x$-$ip_y$ case, the zero-energy states in the isolated vortices acquire finite energies through the band formation. 

\begin{figure}[tb]
  \includegraphics[width=90mm]{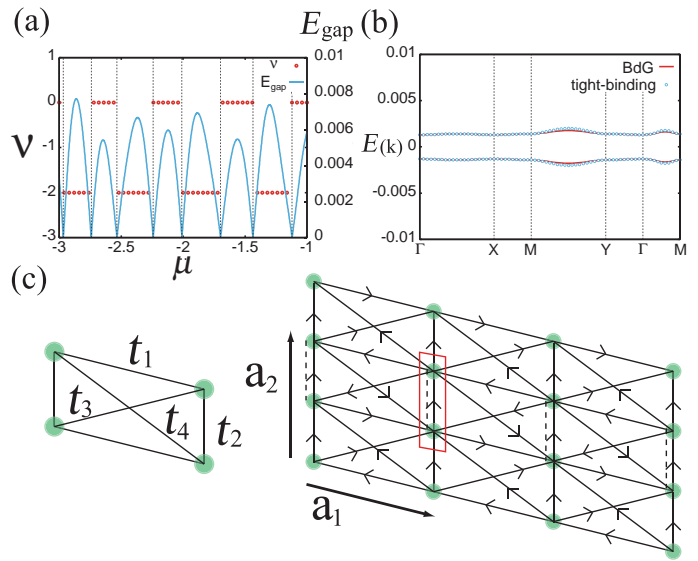}  
  \caption{(a) Energy gap and Chern number for $H/\phi_0=1.25\times 10^{-3}$. Here, the order parameter $\Delta({\mathbf i})$ obtained at $\mu=-2$ is used for all the range of $\mu$.     
    (b) Tight-binding fitting for lowest energy states. We choose $t'=1.41\times10^{-3}$ and $t''=1.31\times10^{-3}$.
    (c) Lattice configuration and transfer integrals for the effective tight-binding model on a general Bravais lattice. A magnetic unit cell is highlighted with
    a red oblique box. The arrow from site $l$ to $m$ indicates $\lambda_{lm}=-\lambda_{ml}=1$. The convention of transfer integrals, $t_1$-$t_4$, and the two lattice vectors, ${\mathbf a}_1$ and ${\mathbf a}_2$, are also depicted.
  }
  \label{fig2}
\end{figure}

While the overall tendency of $E_{\rm gap}$ can be understood as above, however, the field dependence of $E_{\rm gap}$ shows 
non-monotonicity, which cannot be captured in this picture.
In particular, in the $p_x$-$ip_y$ case, $E_{\rm gap}$ shows oscillation [Fig.~\ref{fig1}(f)], and sometimes approaches
zero, 
implying a possibility of quantum phase transition. 
In fact, as shown in Fig.~\ref{fig2}(a), if one sweeps $\mu$, instead of magnetic field, one will find the energy gap closes quite frequently.
Moreover, these successive quantum phase transitions have topological character:
the total Chern number of filled bands, $\nu$, jumps between $0$ and $-2$, every time the gap closes.

The series of topological quantum phase transitions (TQPT) can be well understood in terms of an effective model. Following the procedure in Ref.~\cite{PhysRevB.92.134519}, we fit the lowest two bands by an effective Majorana tight-binding model,
\begin{eqnarray}
{\cal H}_{\rm M}&=it'\sum_{n.n.}\lambda_{lm}\alpha_l\alpha_m + it''\sum_{n.n.n.}\lambda_{lm}\alpha_l\alpha_m,
\label{eq4}
\end{eqnarray}
where $\alpha_l$ is the Majorana operator defined at $l$-th vortex core, $\lambda_{lm}=\pm1$ is the $Z_2$ gauge field, and the summations are taken over pairs of sites $(l, m)$ for nearest- and next-nearest-neighbors in the first and second terms, respectively. 
As shown in Fig.~\ref{fig2}(b), the low-energy bands are well fitted 
by the Hamiltonian Eq.~(\ref{eq4}) at low magnetic fields, by sensitively changing $t'$ and $t''$, as magnetic field or $\mu$.
These effective transfer integrals stem from the overlap integrals of localized modes in the nearby vortex cores, and rapidly oscillates as $\sim\sin(k_{\rm F}R)$,
with Fermi wave vector, $k_{\rm F}$, and inter-vortex distance, $R$.

  Moreover, the Hamiltonian (\ref{eq4}) explains the TQPT discussed above. 
  The Hamiltonian (\ref{eq4}) belongs to the class D~\cite{PRB.78.195125}, and its topological characters
are classified with Chern number, $\nu_{\rm M}$. 
By varying $t'$ and $t''$, the $\nu_{\rm M}$ of lower band changes as
$\nu_{\rm M}=1$ (-1) for $t''>0$ ($t''<0$), 
while the gap vanishes between the bands, at the time-reversal symmetric point, $t''=0$. 
This change of $\nu_{\rm M}$ as $t''$, combined with the sensitive change of transfer integrals with magnetic field or $\mu$,
explains the frequent occurrence of TQPT:
the topological transition of Majorana bands controls the change of total Chern number, $\nu$.
    
Indeed, the TQPT of Majorana bands is a universal feature of 2D chiral superconductors in the vortex state,
which is not related to the details of the system, e.g. the high spatial symmetry of square vortex lattice.
In fact, for a Bravais lattice with general lattice vectors ${\mathbf a}_1$ and ${\mathbf a}_2$, the effective tight-binding model of the type (\ref{eq4}) can be cast into the Hamiltonian analogous to Qi-Hughes-Zhang (QHZ) type\ \cite{PhysRevB.78.195424}:
\begin{eqnarray}
\mathcal{H} = 4\sum_{{\mathbf k},ss'}\alpha^{\dag}_{{\mathbf k}s}{\mathbf d}({\mathbf k})\cdot\bm{\sigma}_{ss'}\alpha_{{\mathbf k}s'},
\label{eq5}
\end{eqnarray}
with $d_1=t_2\sin\frac{k_2}{2}$, $d_2=t_3\cos(k_1+\frac{k_2}{2}) + t_4\cos(k_1-\frac{k_2}{2})$ and $d_3=t_1\sin k_1$, where $k_j={\mathbf k}\cdot{\mathbf a}_j$. 
Here, we assumed only short-range hoppings, $t_1$ to $t_4$ [Fig.~\ref{fig2}(c)].
If one sets $t_1=t_2=t'$ and $t_3=t_4=t''$, the result of square lattice Hamitonian is recovered.
The QHZ Hamiltonian is the simplest model to describe topological phase transition.
Indeed, the Hamiltonian (\ref{eq5}) leads to quantum phase transitions occur in the simple conditions: $t_1=0$ or $t_2=0$ or $t_3+t_4=0$. 

The ubiquitous TQPT gives an important implication to the topological nature of 3D chiral superconductors.
Topological phenomena in different dimensions can often be related with each other through the dimensional reduction.
Here, in order to apply this idea to the vortex state of quasi-2D chiral superconductors, we reconsider the starting Hamiltonian (\ref{eq1}) on a layered square lattice, and introduce small hopping $t_z$, in ${\mathbf z}\ (\parallel{\mathbf H})$ direction. 

\begin{figure}[tb]
\includegraphics[width=80mm]{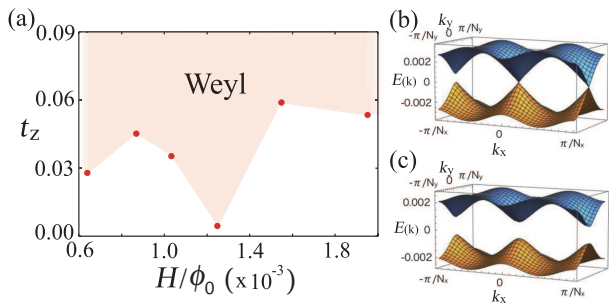}
  \caption{(a) Low-magnetic field phase diagram for the quasi-2D chiral superconductor. The superconducting Weyl phase is realized for larger $t_z$. The $k_z$-projected dispersions are plotted for (b) $k_z=0.415$ and (c) $k_z=2.0$, obtained at $H/\phi_0=1.25\times10^{-3}$ and $t_z=0.05$.}  
  \label{fig3}
\end{figure}

This new setting turns out to make a slight difference in the BdG equation (\ref{eq3}).
Firstly, one needs to introduce the momentum in ${\mathbf z}$ direction, $k_z$, and impose $k_z$ dependence on all the quantities appearing in eq.~(\ref{eq3}).
Through this change, the important observation is that the $k_z$ dependence of $H_{ij}$ can be absorbed into the replacement of $\mu$
with its $k_z$-dependent counterpart: $\mu'(k_z)=\mu-2t_z\cos(k_z)$.
This is practically the only change due to the weak three-dimensionality, if we assume the $k_z$ dependence of $\Delta_{ij}$ is negligible for small $t_z$.
This simple correspondence between 2D and quasi-2D formulations enables us to interpret the physics in these two systems in a unified language.
In particular, the successive TQPT with sweeping $\mu$ in 2D [Fig.~\ref{fig2}(a)] can now be reinterpreted as successive gap closings along $k_z$ axis in momentum space, for the quasi-2D case. In other words, point nodes appear at $k_z$'s,    
corresponding to the quantum critical $\mu=\mu'(k_z)$.
Moreover, Chern number in each $k_z$ slice jumps at point nodes.
This means that all the point nodes are specified by the topological Weyl charges.
Consequently, if $t_z$ is large enough to cross the narrowly-spaced quantum critical $\mu$ [Fig.~\ref{fig2}(a)],
the system shows Weyl superconductivity~\cite{PhysRevB.86.054504,JETP.93.66,PhysRevB.86.104509,PhysRevB.89.020509,PhysRevB.92.214504}.
In Fig.~\ref{fig3} we show the phase diagram.
The superconducting Weyl phase is realized for $t_z\sim 0.01\sim 0.1 \ll t$, which is reasonable for quasi-2D system [Fig.~\ref{fig3}(a)].
In fact, at $H/\phi_0=1.25\times10^{-3}$, the Weyl node appears at $k_z=0.415$ and the gap closes [Fig.~\ref{fig3}(b)], while
the gap opens in the other momenta [Fig.~\ref{fig3}(c)].

In conclusion, we have studied the topological aspect of the chiral $p$-wave superconductors in the vortex state,
with the BdG equation combined with a mapping to effective Majorana tight-binding model.
In the 2D case, we revealed the existence of ubiquitous topological quantum phase transitions, which can be attributed
to only a few universal properties of 2D chiral superconductors, namely, the existence of Majorana modes in isolated vortices,
realization of QHZ-type Hamiltonian from the double enlargement of magnetic unit cell due to $\pi$-flux-carrying vortices, and sensitive variation of effective transfer integrals in the scale of Fermi wavelength.
Through the dimensional reduction, we further showed that a superconducting Weyl phase generically exists in the low-field region of
quasi-2D chiral superconductors. These findings are relevant to general chiral superconductors, potentially including Sr$_2$RuO$_4$,
and further exploration of their physical consequences will be awaited. In fact, the transport phenomena associated with the low-field Weyl phase, and especially its quantum anomaly, will be a fascinating theme of study. Effects of lattice dislocation might also be an interesting issue, which is known to affect the topological nature of 
the system, and a vortex lattice gives a controllable stage for its study. We would like to leave these problems for future study.

The authors are grateful to Y.~Higashi, M.~Takahashi and Y.~Yanase for fruitful discussions.
This work was supported by JSPS KAKENHI (Nos. 26400339, 15H05852, 15K13533 and 16H04026). T. Y. is supported by a JSPS Fellowship for Young Scientists.

\bibliography{reference}
\end{document}